\documentclass{llncs}
\usepackage{makeidx}
\usepackage{times,amssymb,mathptm,graphicx,color}
\definecolor{light}{gray}{.75}
\begin{document}
\title{Efficient Synthesis of Linear Reversible Circuits}
\date{\today}
\author{Ketan N. Patel, Igor L. Markov and John P. Hayes}
\institute{University of Michigan,
           Ann Arbor 48109-2122 \\
           {\tt \{knpatel,imarkov,jhayes\}@eecs.umich.edu}}

\maketitle

\begin{abstract}
In this paper we consider circuit synthesis for $n$-wire linear 
reversible circuits using the C-NOT gate library. These circuits are an 
important class of reversible circuits with applications 
to quantum computation. Previous algorithms, based 
on Gaussian elimination and LU-decomposition, yield circuits with 
$O\left(n^2\right)$ gates in the 
worst-case. However, an information theoretic bound suggests that 
it may be possible to reduce this to as few as 
$O\left(n^2/\log\, n\right)$ gates.

We present an algorithm that 
is optimal up to a multiplicative constant, as well as $\Theta(\log\, n)$ 
times faster than previous methods. While our results are primarily 
asymptotic, simulation results show that even for relatively small $n$ our algorithm is faster and yields 
more efficient circuits than the standard method. Generically our algorithm
can be interpreted as a matrix decomposition algorithm, yielding an asymptotically
efficient decomposition of a binary matrix into a product of elementary matrices. 
\end{abstract}


\section{Introduction}
A reversible or information lossless circuit is one that implements a 
bijective function, or loosely, a circuit where the inputs can be recovered 
from the outputs and all output values are achievable.
A major motivation for studying reversible circuits is the emerging field
of quantum computation~\cite{nielsen:qca:00}. A quantum circuit implements a
unitary function, and is therefore reversible. Circuit synthesis for
reversible computations is an active area of  
research~\cite{barenco:egf:95,cybenko:rqc:01,perkowski:agd:01,shende:rlc:02}. The goal
in circuit synthesis is, given a gate library, 
to synthesize an efficient circuit performing a desired computation. In the
quantum context, the individual gates correspond to physical operations
on quantum states called qubits, and therefore reducing the number of gates in the 
synthesis generally leads to a more efficient implementation.

Linear reversible classical circuits form an important sub-class of
quantum circuits, which can be generated by a single type of 
gate called a C-NOT gate~(see Figure~\ref{fig:ex_gate}c). This gate is an important 
primitive for quantum computation because it forms a universal gate set 
when augmented with single qubit rotations~\cite{divincenzo:tbg:95}. Moreover, 
current quantum circuit synthesis algorithms can generate circuits with
strings of C-NOT gates, and therefore more efficient synthesis for
these classical linear reversible sub-circuits would imply more efficient
synthesis for the overall quantum computation.

In this paper we consider the problem of efficiently synthesizing an 
arbitrary linear reversible circuit on $n$ wires using C-NOT gates. 
This problem can be mapped to the problem of row reducing 
a $n\times n$ binary matrix. Until now the best synthesis methods 
have been based on standard row reduction methods such as Gaussian 
elimination and LU-decomposition, which yield circuits 
with $O(n^2)$ gates~\cite{beth:qaa:01}. However, the 
best lower bound leaves open the possibility that synthesis
with as few as $O\left(n^2/\log\, n\right)$ gates in the 
worst case may exist~\cite{shende:rlc:02}.

We present a new synthesis algorithm that meets the lower bound, 
and is therefore asymptotically optimal up to a multiplicative constant.
Furthermore, our algorithm is also asymptotically faster than 
previous methods. Simulation results show that the proposed algorithm
outperforms previous methods even for relatively small $n$.
Generically our algorithm
can be interpreted as a matrix decomposition algorithm, that yields an asymptotically
efficient elementary matrix decomposition of a binary matrix.
Generalizations to matrices over larger finite fields are straightforward.
\section{Background}
\begin{figure*}[t!]
     \begin{center}
       \includegraphics{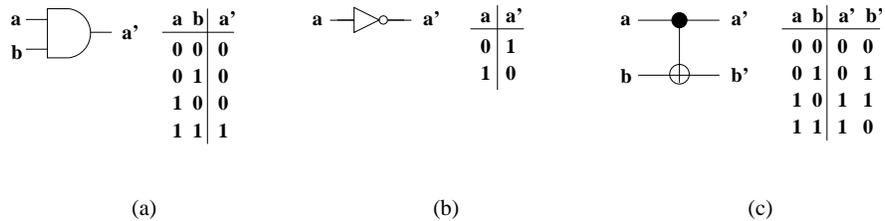} \\
       \caption{Examples of reversible and irreversible logic gates with truth tables a) AND gate\hspace{.03in}  b) NOT gate \hspace{.03in} c) C-NOT gate. Both the NOT and C-NOT gates are reversible while the AND gate is not.} \label{fig:ex_gate}
     \end{center}
  \end{figure*}

We can represent the action of an $n$-input $m$-output logic gate as a 
function mapping the values of
the inputs to those of the outputs: $f: \mathbb{F}_2^n\rightarrow \mathbb{F}_2^m$, where 
$f$ maps each element of $\mathbb{F}_2^n$ to an element in $\mathbb{F}_2^m$. Here $\mathbb{F}_2$ is the 
two-element field, and $\mathbb{F}_2^n$ is the set of all $n$-dimensional vectors over 
this field. A gate is 
\textit{reversible} if this function is bijective, that is, $f$ is one-to-one 
and onto. Intuitively, this means that the inputs 
can be uniquely determined from the outputs and all output values are achievable. 
For example, the AND gate (Figure~\ref{fig:ex_gate}a) is not reversible since 
it maps three input values to the same output value. The NOT gate 
(Figure~\ref{fig:ex_gate}b), on the other hand, is reversible since both possible
input values yield unique output values, and both possible output values are 
achievable. 
The \textit{controlled-NOT} or C-NOT gate, shown in Figure~\ref{fig:ex_gate}c,
is another important reversible gate.
This gate passes the first input, called the \textit{control}, through unchanged 
and inverts the second, called the \textit{target}, if the control is a one. 
As its truth table shows, this gate is reversible since it 
maps each input vector to a unique output vector and all output vectors are 
achievable.

\begin{figure*}[t!]
     \begin{center}
       \includegraphics{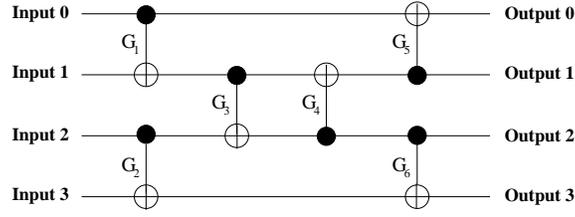} \\
       \caption{Reversible circuit example.} \label{fig:ex_circ}
     \end{center}
  \end{figure*}

A \textit{reversible circuit} is a directed acyclic combinatorial logic circuit where 
all gates are reversible and are interconnected without fanout~\cite{shende:rlc:02}.
An example of a reversible circuit consisting of C-NOT gates is shown
in Figure~\ref{fig:ex_circ}. Note that, as is the case for reversible gates, 
the function computed by a reversible
circuit is bijective.
 
We say a circuit or gate, computing the function $f$, is \textit{linear} if 
$f(x_1 \oplus x_2)=f(x_1)\oplus f(x_2)$ for all $x_1, x_2\in \mathbb{F}_2^n$,
where $\oplus$ is the bitwise XOR operation. The C-NOT gate is an example of a linear
gate:
\[
\begin{array}{ccc}
f\left(\left[0\ 0\right]\right) \oplus f\left(x
    \right)=f\left(x\right)\ \ \  &\ \ \  f\left(\left[0\
    1\right]\right)\oplus f\left(\left[1\
    0\right]\right)= f\left(\left[1\ 1\right]\right)\ \ \ &\ \ \  f\left(\left[1\
    0\right]\right)\oplus f\left(\left[1\
    1\right]\right)= f\left(\left[0\ 1\right]\right) \\
f\left(x\right)\oplus f\left(x\right)=f\left([0\ 0]\right)\ \ \ & \ \ \ f\left(\left[0\ 1\right]\right)\oplus f\left(\left[1 \
    1\right]\right)=f\left(\left[1\ 0\right]\right).\ \ \   
\end{array}
\]
The action of any linear reversible circuit on $n$ wires can be represented by
a linear transformation over $\mathbb{F}_2$. In particular, we can 
represent the action of the circuit as multiplication by a non-singular 
$n\times n$ matrix $A$ with elements in $\mathbb{F}_2$:
\[Ax=y,\]
where $x$ and $y$ are $n$-dimensional vectors representing the values of the input and output
bits respectively. Using this representation, the action of a C-NOT gate 
corresponds to multiplication by an \textit{elementary matrix}, which is the identity matrix with one 
off-diagonal entry set to one. Multiplication by an elementary matrix performs a 
\textit{row operation}, the addition of one row of a matrix or vector to another. 
Applying a series
of C-NOT gates corresponds to performing a series of these row operations on the input 
vector or equivalently to multiplying it by a series of elementary matrices.
For example, the linear transform computed by the 
circuit in 
Figure~\ref{fig:ex_circ} is given by
\[A = \stackrel{G_6}{\left[\begin{array}{cccc}1&0&0&0\\0&1&0&0\\0&0&1&0
\\0&0&1&1\end{array}\right]}\cdot
    \stackrel{G_5}{\left[\begin{array}{cccc}1&1&0&0\\0&1&0&0\\0&0&1&0\\0&0&0&1\end{array}\right]}\cdot
    \stackrel{G_4}{\left[\begin{array}{cccc}1&0&0&0\\0&1&1&0\\0&0&1&0\\0&0&0&1\end{array}\right]}\cdot
    \stackrel{G_3}{\left[\begin{array}{cccc}1&0&0&0\\0&1&0&0\\0&1&1&0\\0&0&0&1\end{array}\right]}\cdot
    \stackrel{G_2}{\left[\begin{array}{cccc}1&0&0&0\\0&1&0&0\\0&0&1&0\\0&0&1&1\end{array}\right]}\cdot
    \stackrel{G_1}{\left[\begin{array}{cccc}1&0&0&0\\1&1&0&0\\0&0&1&0\\0&0&0&1\end{array}\right]}=
    \left[\begin{array}{cccc}1&0&1&0\\0&0&1&0\\1&1&1&0\\1&1&0&1\end{array}\right]\]

We can use the matrix notation to count the number of different $n$-input
linear reversible transformations. In order for the transformation to
be reversible, its matrix must be non-singular, in other words, all
nontrivial sum of the rows should be non-zero. There are $2^n-1$
possible choices for the first row, all vectors except for the all zeros
vector. There are $2^n-2$ possible choices for the second row, since 
it cannot be the equal to the first row or the all zeros vector. In
general, there are $2^n-2^{i-1}$ possible choices for the $i$th row, 
since it cannot be any of the $2^{i-1}$ linear combinations of the
previous $i-1$ rows (otherwise the matrix would be
singular). Therefore there are 
\[\prod_{i=0}^{n-1}\left(2^n-2^{i}\right)\]
unique $n$-input linear reversible transformations.

Since any non-singular matrix $A$ can be
reduced to the identity matrix using row operations, we can write $A$ as 
a product of elementary matrices.
Therefore, any linear reversible function can be be synthesized
from C-NOT gates. Moreover, the problem of C-NOT circuit synthesis
is equivalent to the problem of row reduction of a matrix $A$ 
representing the linear reversible function: any 
synthesis of the circuit can be written as a product of elementary 
matrices equal to $A$ and any such product yields a synthesis. 
The length
of the synthesized circuit is given by the number of elementary matrices
in the product.
Standard Gaussian elimination and LU-decomposition based methods requires 
$O(n^2)$ gates in the worst-case~\cite{beth:qaa:01}. However, the best lower 
bound is only $\Omega\left(n^2/\log\, n\right)$ gates~\cite{shende:rlc:02}.

\begin{lemma}[Lower Bound] \label{lemma:lwr_bound}
There are $n$-bit linear reversible transformation that cannot be synthesized 
using fewer than $\Omega(n^2/\log\, n)$ C-NOT gates.
\end{lemma}
\textbf{Proof}
Let $d$ be the maximum number of C-NOT gates needed to synthesize any 
linear reversible function on $n$ wires.  The number of different 
C-NOT gates which can act on $n$ wires is $n(n-1)$. Therefore the number 
of unique C-NOT circuit with no more than $d$ gates must be no more than 
$\left(n^2-n+1\right)^d$, where we have included a do-nothing NOP gate
in addition to the $n^2-n$ C-NOT gates
to account for circuits with fewer than $d$ gates. Since the number of
circuits with no more than $d$ C-NOT gates must be greater than the
number of unique linear reversible function on $n$ wires, we have the
inequality 
\begin{equation}
\left(n^2-n+1\right)^d \ge \prod_{i=0}^{n-1}\left(2^n-2^i\right)\ge 2^{n(n-1)}.
\end{equation}
Taking the $\log$ of both the left and right sides of the equations gives
\begin{equation}\label{eqn:lwr_bound}
d \ge \frac{n(n-1) \log 2}{\log(n^2-n+1)} = \frac{n^2-n}{\log_2\,(n^2-n+1)}=
\Omega\left(\frac{n^2}{\log\, n}\right).
\end{equation}
$\Box$
 
\noindent This lemma suggests a synthesis more efficient than standard 
Gaussian elimination may be possible. The multiplicative constant in 
this lower bound is $1/2$ (assuming logs are taken base 2).

\section{Efficient Synthesis}
In this section we present our synthesis algorithm, which 
achieves the lower bound given in the previous section. In
Gaussian elimination, row operations are used to 
place ones on the diagonal of the matrix and to eliminate any remaining ones.
One row operation is required for each entry in the matrix that is targeted. 
Since there 
are $n^2$ matrix entries, $O(n^2)$ row operation are required in the worst case. If instead
we group entries together and use single row operations to 
change these groups, we can reduce the number of row operation required, and 
therefore the number of gates needed to synthesize the circuit.

The basic idea is as follows. We first partition the columns of the 
$n\times n$ matrix into sections of no more than $m$ columns each. We call
the entries in a particular row and section a \textit{sub-row}. For
each section we use row operations to eliminate sub-row patterns 
that repeat in that section. This leaves relatively few ($<2^m$) non-zero
sub-rows in the section. These remaining entries are handled using Gaussian 
elimination. If $m$ is small enough ($< \log_2\,n$), most of the row operations
result from the first step, which requires a factor of $m$ fewer row operations
than full Gaussian elimination. As with the Gaussian elimination based method, 
our algorithm is applied in two steps; first the matrix is reduced to an upper 
triangular matrix, the resulting matrix is transposed, and then the process is 
repeated to reduce it to the identity.
Detailed pseudo-code for our algorithm is shown on the next page.  
\begin{figure*}
\begin{center}
\textbf{Algorithm 1: Efficient C-NOT Synthesis}
{\ttfamily \small
  \begin{tabular}{|llllllll|}
  \hline &&&&&&&\\
  &\multicolumn{7}{l|}{\textbf{[circuit] = CNOT\_Synth(A, n, m)}}\\
  &\{$\ $&\multicolumn{6}{l|}{}\\
  &&\multicolumn{6}{l|}{\underline{\rmfamily// synthesize lower/upper triangular part}}\\
  &&\multicolumn{6}{l|}{[A,circuit\_l] = Lwr\_CNOT\_Synth(A, n, m)}\\
  &&\multicolumn{6}{l|}{A = transpose(A);}\\ 
  &&\multicolumn{6}{l|}{[A,circuit\_u] = Lwr\_CNOT\_Synth(A, n, m)}\\
  \multicolumn{8}{|l|}{}\\  
  &&\multicolumn{6}{l|}{\underline{\rmfamily// combine lower/upper triangular synthesis}}\\
  &&\multicolumn{6}{l|}{switch control/target of C-NOT gates in circuit\_u;}\\
  &&\multicolumn{6}{l|}{circuit = [reverse(circuit\_u) | circuit\_l];}\\ 
  &\}&\multicolumn{6}{l|}{}\\
  &&&&&&&\\\hline
  &&&&&&&\\
  &\multicolumn{7}{l|}{\textbf{[A,circuit] = Lwr\_CNOT\_Synth(A, n, m)}}\\
  &\{$\ $&\multicolumn{6}{l|}{}\\
  &&\multicolumn{6}{l|}{circuit = [];}\\
  &&\multicolumn{6}{l|}{for (sec=1; sec<=ceil(n/m); sec++)  \rmfamily// Iterate over column sections}\\
  &&\{$\ $&\multicolumn{5}{l|}{}\\
  &&&\multicolumn{5}{l|}{\underline{\rmfamily// remove duplicate sub-rows in section sec}}\\
  &&&\multicolumn{5}{l|}{for (i=0; i<$2^m$; i++)}\\
  &&&&\multicolumn{4}{l|}{patt[i] = -1; \rmfamily//marker for first positions of sub-row patterns}\\
  &&&\multicolumn{5}{l|}{for (row\_ind=(sec-1)*m; row\_ind<n; row\_ind++)}\\
  &&&\{$\ $&\multicolumn{4}{l|}{}\\
  &&&&\multicolumn{4}{l|}{sub-row\_patt = A[row\_ind,(sec-1)*m:sec*m-1];}\\
  &&&&\multicolumn{4}{l|}{\underline{\rmfamily// if first copy of pattern save otherwise remove}}\\
  &&&&\multicolumn{4}{l|}{if (patt[sub-row\_patt] == -1)}\\
  &&&&&\multicolumn{3}{l|}{patt[sub-row\_patt] = row\_ind;}\\
  &&&&\multicolumn{4}{l|}{else}\\
  &&&&\{$\ $&&&\\
  &&&&&\multicolumn{3}{l|}{A[row\_ind,:] += A[patt[sub-row\_patt],:];}\\
  \textbf{Step A}&&&&&\multicolumn{3}{l|}{circuit = [C-NOT(patt[sub-row\_patt],row\_ind) | circuit];}\\  
  &&&&\}&&&\\
  &&&\}&\multicolumn{4}{l|}{}\\
  &&&\multicolumn{5}{l|}{\underline{\rmfamily// use Gaussian elimination for remaining entries in column section}}\\
  &&&\multicolumn{5}{l|}{for (col\_ind=(sec-1)*m; col\_ind<sec*m-1; col\_ind++)}\\
  &&&\{$\ $&\multicolumn{4}{l|}{}\\
  &&&&\multicolumn{4}{l|}{\underline{\rmfamily// check for 1 on diagonal}}\\
  &&&&\multicolumn{4}{l|}{diag\_one = 1;}\\
  &&&&\multicolumn{4}{l|}{if (A[col\_ind,col\_ind] == 0)}\\
  &&&&&\multicolumn{3}{l|}{diag\_one = 0;}\\
  &&&&\multicolumn{4}{l|}{\underline{\rmfamily// remove ones in rows below col\_ind}}\\  
  &&&&\multicolumn{4}{l|}{for (row\_ind=col\_ind+1; row\_ind<n; row\_ind++)}\\
  &&&&\{$\ $&\multicolumn{3}{l|}{}\\
  &&&&&\multicolumn{3}{l|}{if (A[row\_ind,col\_ind] == 1)}\\
  &&&&&\{$\ $&\multicolumn{2}{l|}{}\\
  &&&&&&\multicolumn{2}{l|}{if (diag\_one == 0)}\\
  &&&&&&\{$\ $&\multicolumn{1}{l|}{}\\
  &&&&&&&\multicolumn{1}{l|}{A[col\_ind,:] += A[row\_ind,:];}\\
  \textbf{Step B}&&&&&&&\multicolumn{1}{l|}{circuit = [C-NOT(row\_ind,col\_ind) | circuit];}\\
  &&&&&&&\multicolumn{1}{l|}{diag\_one = 1;}\\
  &&&&&&\}&\multicolumn{1}{l|}{}\\
  &&&&&&\multicolumn{2}{l|}{A[row\_ind,:] += A[col\_ind,:];}\\
  \textbf{Step C}&&&&&&\multicolumn{2}{l|}{circuit = [C-NOT(col\_ind,row\_ind) | circuit];}\\
  &\}&\}&\}&\}&\}&\multicolumn{2}{l|}{}\\
  &&&&&&&\\
  \hline
  \end{tabular}
} 
\end{center}
\end{figure*}
The following example illustrates our algorithm for a $6$-wire linear
reversible circuit.

\noindent\textbf{1)} Choose $m = 2$ and partition matrix.
\renewcommand{\arraystretch}{.75}
\[\left [\begin {array}{c|c|c|c}
\mbox{\framebox{$\begin{array}{cc}
    1&1\\
    1&0 \end{array}$}} &
   \begin{array}{cc}
    0&0\\
    0&1 \end{array} &
   \begin{array}{cc}
    0&0\\
    1&0 \end{array}\\ 

  \begin{array}{cc}
    0&1\\
    1&1 \end{array} &
\mbox{\framebox{$\begin{array}{cc}
    0&0\\
    1&1 \end{array}$}} &
   \begin{array}{cc}
    1&0\\
    1&1 \end{array} \\

  \begin{array}{cc}
    1&1\\
    0&0 \end{array} &
  \begin{array}{cc}
    0&1\\
    1&1 \end{array} &
\mbox{\framebox{$\begin{array}{cc}
    1&1\\
    1&0 \end{array}$}} 

\end{array}\right]\]
\noindent\textbf{2)} (Step A - section 1) Eliminate duplicate sub-rows.
\[\left [\begin {array}{c|c|c|c}
\mbox{\framebox{$\begin{array}{cc}
    1&1\\
    1&0 \end{array}$}} &
   \begin{array}{cc}
    0&0\\
    0&1 \end{array} &
   \begin{array}{cc}
    0&0\\
    1&0 \end{array}\\ 

  \begin{array}{cc}
    0&1\\
    \colorbox{light}{1}&\colorbox{light}{1} \end{array} &
\mbox{\framebox{$\begin{array}{cc}
    0&0\\
    1&1 \end{array}$}} &
   \begin{array}{cc}
    1&0\\
    1&1 \end{array} \\

  \begin{array}{cc}
    \colorbox{light}{1}&\colorbox{light}{1}\\
    0&0 \end{array} &
  \begin{array}{cc}
    0&1\\
    1&1 \end{array} &
\mbox{\framebox{$\begin{array}{cc}
    1&1\\
    1&0 \end{array}$}} 

\end{array}\right]\stackrel{\begin{array}{c} 1\rightarrow 4\\ 1\rightarrow 5\end{array}}{\Longrightarrow}
\left [\begin {array}{c|c|c|c}
\mbox{\framebox{$\begin{array}{cc}
    1&1\\
    1&0 \end{array}$}} &
   \begin{array}{cc}
    0&0\\
    0&1 \end{array} &
   \begin{array}{cc}
    0&0\\
    1&0 \end{array}\\ 

  \begin{array}{cc}
    0&1\\
    0&0 \end{array} &
\mbox{\framebox{$\begin{array}{cc}
    0&0\\
    1&1 \end{array}$}} &
   \begin{array}{cc}
    1&0\\
    1&1 \end{array} \\

  \begin{array}{cc}
    0&0\\
    0&0 \end{array} &
  \begin{array}{cc}
    0&1\\
    1&1 \end{array} &
\mbox{\framebox{$\begin{array}{cc}
    1&1\\
    1&0 \end{array}$}} 

\end{array}\right]
\]
\noindent\textbf{3)} (Step B - section 1, column 1) One already on diagonal.

\noindent\textbf{4)} (Step C - section 1, column 1) Remove remaining ones in column below diagonal.
\[\left [\begin {array}{c|c|c|c}
\mbox{\framebox{$\begin{array}{cc}
    1&1\\
    \colorbox{light}{1}&0 \end{array}$}} &
   \begin{array}{cc}
    0&0\\
    0&1 \end{array} &
   \begin{array}{cc}
    0&0\\
    1&0 \end{array}\\ 

  \begin{array}{cc}
    0&1\\
    0&0 \end{array} &
\mbox{\framebox{$\begin{array}{cc}
    0&0\\
    1&1 \end{array}$}} &
   \begin{array}{cc}
    1&0\\
    1&1 \end{array} \\

  \begin{array}{cc}
    0&0\\
    0&0 \end{array} &
  \begin{array}{cc}
    0&1\\
    1&1 \end{array} &
\mbox{\framebox{$\begin{array}{cc}
    1&1\\
    1&0 \end{array}$}} 

\end{array}\right]\stackrel{\begin{array}{c} 1\rightarrow 2\end{array}}{\Longrightarrow}
\left [\begin {array}{c|c|c|c}
\mbox{\framebox{$\begin{array}{cc}
    1&1\\
    0&1 \end{array}$}} &
   \begin{array}{cc}
    0&0\\
    0&1 \end{array} &
   \begin{array}{cc}
    0&0\\
    1&0 \end{array}\\ 

  \begin{array}{cc}
    0&1\\
    0&0 \end{array} &
\mbox{\framebox{$\begin{array}{cc}
    0&0\\
    1&1 \end{array}$}} &
   \begin{array}{cc}
    1&0\\
    1&1 \end{array} \\

  \begin{array}{cc}
    0&0\\
    0&0 \end{array} &
  \begin{array}{cc}
    0&1\\
    1&1 \end{array} &
\mbox{\framebox{$\begin{array}{cc}
    1&1\\
    1&0 \end{array}$}} 

\end{array}\right]
\]
\noindent\textbf{3)} (Step B - section 1, column 1) One already on diagonal.

\noindent\textbf{4)} (Step C - section 1, column 1) Remove remaining ones in column below diagonal.
\[\left [\begin {array}{c|c|c|c}
\mbox{\framebox{$\begin{array}{cc}
    1&1\\
    0&1 \end{array}$}} &
   \begin{array}{cc}
    0&0\\
    0&1 \end{array} &
   \begin{array}{cc}
    0&0\\
    1&0 \end{array}\\ 

  \begin{array}{cc}
    0&\colorbox{light}{1}\\
    0&0 \end{array} &
\mbox{\framebox{$\begin{array}{cc}
    0&0\\
    1&1 \end{array}$}} &
   \begin{array}{cc}
    1&0\\
    1&1 \end{array} \\

  \begin{array}{cc}
    0&0\\
    0&0 \end{array} &
  \begin{array}{cc}
    0&1\\
    1&1 \end{array} &
\mbox{\framebox{$\begin{array}{cc}
    1&1\\
    1&0 \end{array}$}} 

\end{array}\right]\stackrel{\begin{array}{c} 2\rightarrow 3\end{array}}{\Longrightarrow}
\left [\begin {array}{c|c|c|c}
\mbox{\framebox{$\begin{array}{cc}
    1&1\\
    0&1 \end{array}$}} &
   \begin{array}{cc}
    0&0\\
    0&1 \end{array} &
   \begin{array}{cc}
    0&0\\
    1&0 \end{array}\\ 

  \begin{array}{cc}
    0&0\\
    0&0 \end{array} &
\mbox{\framebox{$\begin{array}{cc}
    0&1\\
    1&1 \end{array}$}} &
   \begin{array}{cc}
    0&0\\
    1&1 \end{array} \\

  \begin{array}{cc}
    0&0\\
    0&0 \end{array} &
  \begin{array}{cc}
    0&1\\
    1&1 \end{array} &
\mbox{\framebox{$\begin{array}{cc}
    1&1\\
    1&0 \end{array}$}} 

\end{array}\right]
\]
\noindent\textbf{5)} (Step A - section 2) Eliminate duplicate sub-rows below row 2.
\[\left [\begin {array}{c|c|c|c}
\mbox{\framebox{$\begin{array}{cc}
    1&1\\
    0&1 \end{array}$}} &
   \begin{array}{cc}
    0&0\\
    0&1 \end{array} &
   \begin{array}{cc}
    0&0\\
    1&0 \end{array}\\ 

  \begin{array}{cc}
    0&0\\
    0&0 \end{array} &
\mbox{\framebox{$\begin{array}{cc}
    0&1\\
    1&1 \end{array}$}} &
   \begin{array}{cc}
    0&0\\
    1&1 \end{array} \\

  \begin{array}{cc}
    0&0\\
    0&0 \end{array} &
  \begin{array}{cc}
    \colorbox{light}{0}&\colorbox{light}{1}\\
    \colorbox{light}{1}&\colorbox{light}{1} \end{array} &
\mbox{\framebox{$\begin{array}{cc}
    1&1\\
    1&0 \end{array}$}} 
\end{array}\right]\stackrel{\begin{array}{c} 3\rightarrow 5\\4\rightarrow 6\end{array}}{\Longrightarrow}
\left [\begin {array}{c|c|c|c}
\mbox{\framebox{$\begin{array}{cc}
    1&1\\
    0&1 \end{array}$}} &
   \begin{array}{cc}
    0&0\\
    0&1 \end{array} &
   \begin{array}{cc}
    0&0\\
    1&0 \end{array}\\ 

  \begin{array}{cc}
    0&0\\
    0&0 \end{array} &
\mbox{\framebox{$\begin{array}{cc}
    0&1\\
    1&1 \end{array}$}} &
   \begin{array}{cc}
    0&0\\
    1&1 \end{array} \\

  \begin{array}{cc}
    0&0\\
    0&0 \end{array} &
  \begin{array}{cc}
    0&0\\
    0&0 \end{array} &
\mbox{\framebox{$\begin{array}{cc}
    1&1\\
    0&1 \end{array}$}} 
\end{array}\right]
\]
\noindent\textbf{6)} (Step B - section 2, column 3) Place one on diagonal.
\[\left [\begin {array}{c|c|c|c}
\mbox{\framebox{$\begin{array}{cc}
    1&1\\
    0&1 \end{array}$}} &
   \begin{array}{cc}
    0&0\\
    0&1 \end{array} &
   \begin{array}{cc}
    0&0\\
    1&0 \end{array}\\ 

  \begin{array}{cc}
    0&0\\
    0&0 \end{array} &
\mbox{\framebox{$\begin{array}{cc}
    \colorbox{light}{0}&1\\
    1&1 \end{array}$}} &
   \begin{array}{cc}
    0&0\\
    1&1 \end{array} \\

  \begin{array}{cc}
    0&0\\
    0&0 \end{array} &
  \begin{array}{cc}
    0&0\\
    0&0 \end{array} &
\mbox{\framebox{$\begin{array}{cc}
    1&1\\
    0&1 \end{array}$}} 
\end{array}\right]\stackrel{\begin{array}{c} 4\rightarrow 3\end{array}}{\Longrightarrow}
\left [\begin {array}{c|c|c|c}
\mbox{\framebox{$\begin{array}{cc}
    1&1\\
    0&1 \end{array}$}} &
   \begin{array}{cc}
    0&0\\
    0&1 \end{array} &
   \begin{array}{cc}
    0&0\\
    1&0 \end{array}\\ 

  \begin{array}{cc}
    0&0\\
    0&0 \end{array} &
\mbox{\framebox{$\begin{array}{cc}
    1&0\\
    1&1 \end{array}$}} &
   \begin{array}{cc}
    1&1\\
    1&1 \end{array} \\

  \begin{array}{cc}
    0&0\\
    0&0 \end{array} &
  \begin{array}{cc}
    0&0\\
    0&0 \end{array} &
\mbox{\framebox{$\begin{array}{cc}
    1&1\\
    0&1 \end{array}$}} 
\end{array}\right]
\]
\noindent\textbf{7)} (Step C - section 2, column 3) Remove remaining ones in column below diagonal.
\[\left [\begin {array}{c|c|c|c}
\mbox{\framebox{$\begin{array}{cc}
    1&1\\
    0&1 \end{array}$}} &
   \begin{array}{cc}
    0&0\\
    0&1 \end{array} &
   \begin{array}{cc}
    0&0\\
    1&0 \end{array}\\ 

  \begin{array}{cc}
    0&0\\
    0&0 \end{array} &
\mbox{\framebox{$\begin{array}{cc}
    1&0\\
    \colorbox{light}{1}&1 \end{array}$}} &
   \begin{array}{cc}
    1&1\\
    1&1 \end{array} \\

  \begin{array}{cc}
    0&0\\
    0&0 \end{array} &
  \begin{array}{cc}
    0&0\\
    0&0 \end{array} &
\mbox{\framebox{$\begin{array}{cc}
    1&1\\
    0&1 \end{array}$}} 
\end{array}\right]\stackrel{\begin{array}{c} 3\rightarrow 4\end{array}}{\Longrightarrow}
\left [\begin {array}{c|c|c|c}
\mbox{\framebox{$\begin{array}{cc}
    1&1\\
    0&1 \end{array}$}} &
   \begin{array}{cc}
    0&0\\
    0&1 \end{array} &
   \begin{array}{cc}
    0&0\\
    1&0 \end{array}\\ 

  \begin{array}{cc}
    0&0\\
    0&0 \end{array} &
\mbox{\framebox{$\begin{array}{cc}
    1&0\\
    0&1 \end{array}$}} &
   \begin{array}{cc}
    1&1\\
    0&0 \end{array} \\

  \begin{array}{cc}
    0&0\\
    0&0 \end{array} &
  \begin{array}{cc}
    0&0\\
    0&0 \end{array} &
\mbox{\framebox{$\begin{array}{cc}
    1&1\\
    0&1 \end{array}$}} 
\end{array}\right]
\]
\noindent\textbf{8)} Matrix is now upper triangular. Transpose and continue.
\[\left [\begin {array}{c|c|c|c}
\mbox{\framebox{$\begin{array}{cc}
    1&1\\
    0&1 \end{array}$}} &
   \begin{array}{cc}
    0&0\\
    0&1 \end{array} &
   \begin{array}{cc}
    0&0\\
    1&0 \end{array}\\ 

  \begin{array}{cc}
    0&0\\
    0&0 \end{array} &
\mbox{\framebox{$\begin{array}{cc}
    1&0\\
    0&1 \end{array}$}} &
   \begin{array}{cc}
    1&1\\
    0&0 \end{array} \\

  \begin{array}{cc}
    0&0\\
    0&0 \end{array} &
  \begin{array}{cc}
    0&0\\
    0&0 \end{array} &
\mbox{\framebox{$\begin{array}{cc}
    1&1\\
    0&1 \end{array}$}} 
\end{array}\right]\stackrel{\begin{array}{c} \mbox{transpose}\end{array}}{\Longrightarrow}
\left [\begin {array}{c|c|c|c}
\mbox{\framebox{$\begin{array}{cc}
    1&0\\
    1&1 \end{array}$}} &
   \begin{array}{cc}
    0&0\\
    0&0 \end{array} &
   \begin{array}{cc}
    0&0\\
    0&0 \end{array}\\ 

  \begin{array}{cc}
    0&0\\
    0&1 \end{array} &
\mbox{\framebox{$\begin{array}{cc}
    1&0\\
    0&1 \end{array}$}} &
   \begin{array}{cc}
    0&0\\
    0&0 \end{array} \\

  \begin{array}{cc}
    0&1\\
    0&0 \end{array} &
  \begin{array}{cc}
    1&0\\
    1&0 \end{array} &
\mbox{\framebox{$\begin{array}{cc}
    1&0\\
    1&1 \end{array}$}} 
\end{array}\right]
\]
\noindent\textbf{9)} (Step A - section 1) Eliminate duplicate sub-rows.
\[\left [\begin {array}{c|c|c|c}
\mbox{\framebox{$\begin{array}{cc}
    1&0\\
    1&1 \end{array}$}} &
   \begin{array}{cc}
    0&0\\
    0&0 \end{array} &
   \begin{array}{cc}
    0&0\\
    0&0 \end{array}\\ 

  \begin{array}{cc}
    0&0\\
    0&1 \end{array} &
\mbox{\framebox{$\begin{array}{cc}
    1&0\\
    0&1 \end{array}$}} &
   \begin{array}{cc}
    0&0\\
    0&0 \end{array} \\

  \begin{array}{cc}
    \colorbox{light}{0}&\colorbox{light}{1}\\
    0&0 \end{array} &
  \begin{array}{cc}
    1&0\\
    1&0 \end{array} &
\mbox{\framebox{$\begin{array}{cc}
    1&0\\
    1&1 \end{array}$}} 
\end{array}\right]\stackrel{\begin{array}{c} 4\rightarrow 5\end{array}}{\Longrightarrow}
\left [\begin {array}{c|c|c|c}
\mbox{\framebox{$\begin{array}{cc}
    1&0\\
    1&1 \end{array}$}} &
   \begin{array}{cc}
    0&0\\
    0&0 \end{array} &
   \begin{array}{cc}
    0&0\\
    0&0 \end{array}\\ 

  \begin{array}{cc}
    0&0\\
    0&1 \end{array} &
\mbox{\framebox{$\begin{array}{cc}
    1&0\\
    0&1 \end{array}$}} &
   \begin{array}{cc}
    0&0\\
    0&0 \end{array} \\

  \begin{array}{cc}
    0&0\\
    0&0 \end{array} &
  \begin{array}{cc}
    1&1\\
    1&0 \end{array} &
\mbox{\framebox{$\begin{array}{cc}
    1&0\\
    1&1 \end{array}$}} 
\end{array}\right]
\]
\begin{tabbing}
\noindent\textbf{10)}~\= (Step B - section 1, column 1)  Because matrix is triangular and non-singular there \\
\>will always be ones on the diagonal.
\end{tabbing}
\noindent\textbf{11)} (Step C - section 1, column 1) Remove remaining ones in column.
\[\left [\begin {array}{c|c|c|c}
\mbox{\framebox{$\begin{array}{cc}
    1&0\\
    \colorbox{light}1&1 \end{array}$}} &
   \begin{array}{cc}
    0&0\\
    0&0 \end{array} &
   \begin{array}{cc}
    0&0\\
    0&0 \end{array}\\ 

  \begin{array}{cc}
    0&0\\
    0&1 \end{array} &
\mbox{\framebox{$\begin{array}{cc}
    1&0\\
    0&1 \end{array}$}} &
   \begin{array}{cc}
    0&0\\
    0&0 \end{array} \\

  \begin{array}{cc}
    0&0\\
    0&0 \end{array} &
  \begin{array}{cc}
    1&1\\
    1&0 \end{array} &
\mbox{\framebox{$\begin{array}{cc}
    1&0\\
    1&1 \end{array}$}} 
\end{array}\right]
\stackrel{\begin{array}{c} 1\rightarrow 2\end{array}}{\Longrightarrow}
\left [\begin {array}{c|c|c|c}
\mbox{\framebox{$\begin{array}{cc}
    1&0\\
    0&1 \end{array}$}} &
   \begin{array}{cc}
    0&0\\
    0&0 \end{array} &
   \begin{array}{cc}
    0&0\\
    0&0 \end{array}\\ 

  \begin{array}{cc}
    0&0\\
    0&1 \end{array} &
\mbox{\framebox{$\begin{array}{cc}
    1&0\\
    0&1 \end{array}$}} &
   \begin{array}{cc}
    0&0\\
    0&0 \end{array} \\

  \begin{array}{cc}
    0&0\\
    0&0 \end{array} &
  \begin{array}{cc}
    1&1\\
    1&0 \end{array} &
\mbox{\framebox{$\begin{array}{cc}
    1&0\\
    1&1 \end{array}$}} 
\end{array}\right]
\]
\noindent\textbf{12)} (Step C - section 1, column 2) Remove remaining ones in column.
\[\left [\begin {array}{c|c|c|c}
\mbox{\framebox{$\begin{array}{cc}
    1&0\\
    0&1 \end{array}$}} &
   \begin{array}{cc}
    0&0\\
    0&0 \end{array} &
   \begin{array}{cc}
    0&0\\
    0&0 \end{array}\\ 

  \begin{array}{cc}
    0&0\\
    0&\colorbox{light}{1} \end{array} &
\mbox{\framebox{$\begin{array}{cc}
    1&0\\
    0&1 \end{array}$}} &
   \begin{array}{cc}
    0&0\\
    0&0 \end{array} \\

  \begin{array}{cc}
    0&0\\
    0&0 \end{array} &
  \begin{array}{cc}
    1&1\\
    1&0 \end{array} &
\mbox{\framebox{$\begin{array}{cc}
    1&0\\
    1&1 \end{array}$}} 
\end{array}\right]
\stackrel{\begin{array}{c} 2\rightarrow 4\end{array}}{\Longrightarrow}
\left [\begin {array}{c|c|c|c}
\mbox{\framebox{$\begin{array}{cc}
    1&0\\
    0&1 \end{array}$}} &
   \begin{array}{cc}
    0&0\\
    0&0 \end{array} &
   \begin{array}{cc}
    0&0\\
    0&0 \end{array}\\ 

  \begin{array}{cc}
    0&0\\
    0&0 \end{array} &
\mbox{\framebox{$\begin{array}{cc}
    1&0\\
    0&1 \end{array}$}} &
   \begin{array}{cc}
    0&0\\
    0&0 \end{array} \\

  \begin{array}{cc}
    0&0\\
    0&0 \end{array} &
  \begin{array}{cc}
    1&1\\
    1&0 \end{array} &
\mbox{\framebox{$\begin{array}{cc}
    1&0\\
    1&1 \end{array}$}} 
\end{array}\right]
\]
\noindent\textbf{13)} (Step A - section 2) Eliminate duplicate sub-rows.
\[\left [\begin {array}{c|c|c|c}
\mbox{\framebox{$\begin{array}{cc}
    1&0\\
    0&1 \end{array}$}} &
   \begin{array}{cc}
    0&0\\
    0&0 \end{array} &
   \begin{array}{cc}
    0&0\\
    0&0 \end{array}\\ 

  \begin{array}{cc}
    0&0\\
    0&0 \end{array} &
\mbox{\framebox{$\begin{array}{cc}
    1&0\\
    0&1 \end{array}$}} &
   \begin{array}{cc}
    0&0\\
    0&0 \end{array} \\

  \begin{array}{cc}
    0&0\\
    0&0 \end{array} &
  \begin{array}{cc}
    1&1\\
    \colorbox{light}{1}&\colorbox{light}{0} \end{array} &
\mbox{\framebox{$\begin{array}{cc}
    1&0\\
    1&1 \end{array}$}} 
\end{array}\right]
\stackrel{\begin{array}{c} 3\rightarrow 6\end{array}}{\Longrightarrow}
\left [\begin {array}{c|c|c|c}
\mbox{\framebox{$\begin{array}{cc}
    1&0\\
    0&1 \end{array}$}} &
   \begin{array}{cc}
    0&0\\
    0&0 \end{array} &
   \begin{array}{cc}
    0&0\\
    0&0 \end{array}\\ 

  \begin{array}{cc}
    0&0\\
    0&0 \end{array} &
\mbox{\framebox{$\begin{array}{cc}
    1&0\\
    0&1 \end{array}$}} &
   \begin{array}{cc}
    0&0\\
    0&0 \end{array} \\

  \begin{array}{cc}
    0&0\\
    0&0 \end{array} &
  \begin{array}{cc}
    1&1\\
    0&0 \end{array} &
\mbox{\framebox{$\begin{array}{cc}
    1&0\\
    1&1 \end{array}$}} 
\end{array}\right]
\]
\noindent\textbf{14)} (Step C - section 2, column 1) Remove remaining ones in column.
\[\left [\begin {array}{c|c|c|c}
\mbox{\framebox{$\begin{array}{cc}
    1&0\\
    0&1 \end{array}$}} &
   \begin{array}{cc}
    0&0\\
    0&0 \end{array} &
   \begin{array}{cc}
    0&0\\
    0&0 \end{array}\\ 

  \begin{array}{cc}
    0&0\\
    0&0 \end{array} &
\mbox{\framebox{$\begin{array}{cc}
    1&0\\
    0&1 \end{array}$}} &
   \begin{array}{cc}
    0&0\\
    0&0 \end{array} \\

  \begin{array}{cc}
    0&0\\
    0&0 \end{array} &
  \begin{array}{cc}
    \colorbox{light}{1}&1\\
    0&0 \end{array} &
\mbox{\framebox{$\begin{array}{cc}
    1&0\\
    1&1 \end{array}$}} 
\end{array}\right]
\stackrel{\begin{array}{c} 3\rightarrow 5\end{array}}{\Longrightarrow}
\left [\begin {array}{c|c|c|c}
\mbox{\framebox{$\begin{array}{cc}
    1&0\\
    0&1 \end{array}$}} &
   \begin{array}{cc}
    0&0\\
    0&0 \end{array} &
   \begin{array}{cc}
    0&0\\
    0&0 \end{array}\\ 

  \begin{array}{cc}
    0&0\\
    0&0 \end{array} &
\mbox{\framebox{$\begin{array}{cc}
    1&0\\
    0&1 \end{array}$}} &
   \begin{array}{cc}
    0&0\\
    0&0 \end{array} \\

  \begin{array}{cc}
    0&0\\
    0&0 \end{array} &
  \begin{array}{cc}
    0&1\\
    0&0 \end{array} &
\mbox{\framebox{$\begin{array}{cc}
    1&0\\
    1&1 \end{array}$}} 
\end{array}\right]
\]
\noindent\textbf{15)} (Step C - section 2, column 2) Remove remaining ones in column.
\[\left [\begin {array}{c|c|c|c}
\mbox{\framebox{$\begin{array}{cc}
    1&0\\
    0&1 \end{array}$}} &
   \begin{array}{cc}
    0&0\\
    0&0 \end{array} &
   \begin{array}{cc}
    0&0\\
    0&0 \end{array}\\ 

  \begin{array}{cc}
    0&0\\
    0&0 \end{array} &
\mbox{\framebox{$\begin{array}{cc}
    1&0\\
    0&1 \end{array}$}} &
   \begin{array}{cc}
    0&0\\
    0&0 \end{array} \\

  \begin{array}{cc}
    0&0\\
    0&0 \end{array} &
  \begin{array}{cc}
    0&\colorbox{light}{1}\\
    0&0 \end{array} &
\mbox{\framebox{$\begin{array}{cc}
    1&0\\
    1&1 \end{array}$}} 
\end{array}\right]
\stackrel{\begin{array}{c} 4\rightarrow 5\end{array}}{\Longrightarrow}
\left [\begin {array}{c|c|c|c}
\mbox{\framebox{$\begin{array}{cc}
    1&0\\
    0&1 \end{array}$}} &
   \begin{array}{cc}
    0&0\\
    0&0 \end{array} &
   \begin{array}{cc}
    0&0\\
    0&0 \end{array}\\ 

  \begin{array}{cc}
    0&0\\
    0&0 \end{array} &
\mbox{\framebox{$\begin{array}{cc}
    1&0\\
    0&1 \end{array}$}} &
   \begin{array}{cc}
    0&0\\
    0&0 \end{array} \\

  \begin{array}{cc}
    0&0\\
    0&0 \end{array} &
  \begin{array}{cc}
    0&0\\
    0&0 \end{array} &
\mbox{\framebox{$\begin{array}{cc}
    1&0\\
    1&1 \end{array}$}} 
\end{array}\right]
\]
\noindent\textbf{16)} (Step C - section 3, column 1) Remove remaining ones in column.
\[\left [\begin {array}{c|c|c|c}
\mbox{\framebox{$\begin{array}{cc}
    1&0\\
    0&1 \end{array}$}} &
   \begin{array}{cc}
    0&0\\
    0&0 \end{array} &
   \begin{array}{cc}
    0&0\\
    0&0 \end{array}\\ 

  \begin{array}{cc}
    0&0\\
    0&0 \end{array} &
\mbox{\framebox{$\begin{array}{cc}
    1&0\\
    0&1 \end{array}$}} &
   \begin{array}{cc}
    0&0\\
    0&0 \end{array} \\

  \begin{array}{cc}
    0&0\\
    0&0 \end{array} &
  \begin{array}{cc}
    0&0\\
    0&0 \end{array} &
\mbox{\framebox{$\begin{array}{cc}
    1&0\\
    \colorbox{light}{1}&1 \end{array}$}} 
\end{array}\right]
\stackrel{\begin{array}{c} 5\rightarrow 6\end{array}}{\Longrightarrow}
\left [\begin {array}{c|c|c|c}
\mbox{\framebox{$\begin{array}{cc}
    1&0\\
    0&1 \end{array}$}} &
   \begin{array}{cc}
    0&0\\
    0&0 \end{array} &
   \begin{array}{cc}
    0&0\\
    0&0 \end{array}\\ 

  \begin{array}{cc}
    0&0\\
    0&0 \end{array} &
\mbox{\framebox{$\begin{array}{cc}
    1&0\\
    0&1 \end{array}$}} &
   \begin{array}{cc}
    0&0\\
    0&0 \end{array} \\

  \begin{array}{cc}
    0&0\\
    0&0 \end{array} &
  \begin{array}{cc}
    0&0\\
    0&0 \end{array} &
\mbox{\framebox{$\begin{array}{cc}
    1&0\\
    0&1 \end{array}$}} 
\end{array}\right]
\]
\begin{figure*}[t!]
     \begin{center}
       \includegraphics{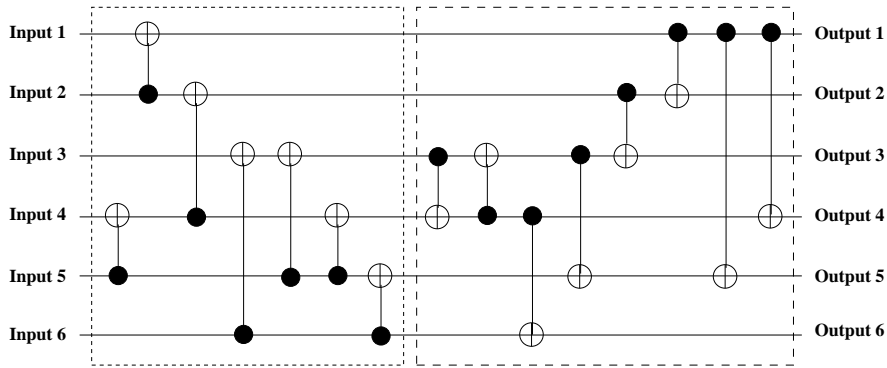} \\
       \caption{Synthesized C-NOT circuit example. The gates in the 
right and left boxes correspond to row operations before and after
the transpose step respectively. Those in the left box are in the 
same order the row operations were applied and their controls and 
targets are switched. The gates in the right box are in the reverse 
order that the row operations were applied.} \label{fig:synth_circ}
     \end{center}
  \end{figure*}
The synthesized circuit is specified by the row operations and is shown in 
Figure~\ref{fig:synth_circ}. 

In general, the length of the synthesized circuit is given by the number of 
row operations
used in the algorithm. By accounting for the maximum number of row operations
in each step, we can calculate an upper bound on the
maximum number of gates that could be required in synthesizing an
$n$-wire linear reversible circuit. C-NOT gates are added in the steps marked
Step A-C in the algorithm. Step A is used to eliminate the duplicates 
in the subsections. It is called fewer than $n+m$ times per section (combined 
for the upper/lower triangular stages of the algorithm), giving 
a total of no more than $(n+m)\cdot \lceil n/m\rceil$ gates. Step B is used
to place ones on the diagonal. It can be called no more than $n$ times.
Step C is used to remove the ones remaining after all duplicate sub-rows
have been cleared. Since there are only $2^m$ $m$-bit words,
there can be at most as many non-zero sub-rows below the $m\times m$ 
sub-matrix on the diagonal. Therefore, Step C is called fewer than $m\cdot (2^m+m)$ times per section, or fewer than
$2\lceil n/m\rceil m\cdot (2^m+m)$ times in all. Adding these up we have
\begin{eqnarray}
\mbox{total row ops} &\le& (n+m)\cdot \left\lceil \frac{n}{m}\right\rceil + n + 2\left\lceil \frac{n}{m}\right\rceil m\cdot \left(2^m+m\right) \\
  &\le&
\frac{n^2}{m} + n + n + m + n + 2n2^m + 2nm + 2m2^m + 2m^2.
\end{eqnarray}
If $m=\alpha \log_2\,n$,
\begin{eqnarray}
\mbox{total row ops} \le \frac{n^2}{\alpha \log_2\, n} &+& 3n + \alpha \log_2\, n+ 2n^{1+\alpha} + 2n\alpha\log_2\,n\nonumber \\
 &+&2\alpha\log_2\,n \cdot n^{\alpha}+2\left(\alpha \log_2\, n\right)^2. \label{eqn:row_ops}
\end{eqnarray}
If $\alpha < 1$, the first term dominates as $n$ gets large. Therefore the number of row operations is $O(n^2/\log\, n)$. Combining this result with Lemma~\ref{lemma:lwr_bound}, we have the following theorem.

\begin{theorem} 
The worst-case length of an $n$-wire C-NOT circuit is $\Theta(n^2/\log\, n)$
gates.
\end{theorem}
In Equation~\ref{eqn:row_ops}, $\alpha$ can be chosen to be arbitrarily close 
to 1. In the limit, the multiplicative constant in the $O(n^2/\log\, n)$ 
expression becomes $1$ (assuming logs are taken base 2). By contrast, the multiplicative 
constant in the lower bound in Lemma~\ref{lemma:lwr_bound} is $1/2$.

This algorithm, in addition to generating more efficient circuits than the 
standard method, is also asymptotically more efficient in terms of run time. 
The execution time of the algorithm is dominated by the row operations on 
the matrix, which are each $O(n)$. Therefore the overall execution time is 
$O(n^3/\log\, n)$ compared to $O(n^3)$ for standard Gaussian 
elimination~\cite[p. 42]{press:nrc:92}. 

Our algorithm 
is closely related to Kronrod's Algorithm~(also known as ``The Four Russians' Algorithm'') 
for construction of the transitive closure of a graph~\cite{arlazarov:oec:70}.   
One important difference between the two is that in their case the goal was 
a fast algorithm for their application, which is only of secondary concern 
for our application. Our primary goal is an algorithm 
that produces an efficient circuit synthesis. Generically, our algorithm can
be interpreted as producing
an efficient elementary matrix decomposition of a binary matrix.

\section{Simulation Results}
\begin{figure*}[t!]
     \begin{center}
       \resizebox{4in}{!}
       {\includegraphics{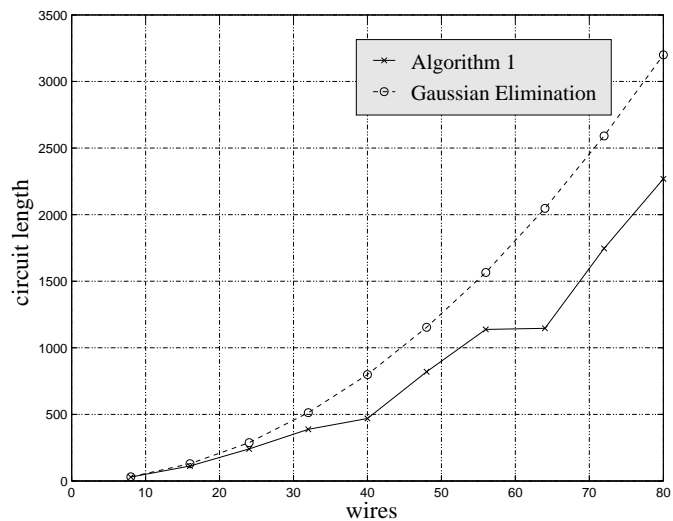}} \\
       \caption{Performance of Algorithm 1 vs. Gaussian elimination on randomly generated linear reversible functions. Each point corresponds to the average length of the circuit generated for 100 randomly generated matrices. The x-axis specifies $n$, the number of inputs/outputs of the linear reversible circuit, and the y-axis specifies the average number of gates in the circuit synthesis. For Algorithm 1, $m$ was chosen to be $\mbox{round}\left((\log_2\,n)/2\right)$.} \label{fig:avg_length}
     \end{center}
  \end{figure*}
Though Algorithm 1 is asymptotically optimal, it would
be of interest to know how large $n$ must be before the algorithm begins to 
outperform standard Gaussian elimination. For this purpose we have synthesized 
linear reversible circuits using both our method and Gaussian elimination for 
randomly generated non-singular 0-1 matrices. The results of these simulations
are summarized in Figure~\ref{fig:avg_length}. Our algorithm shows an
improvement over Gaussian elimination for $n$ as small as $8$. The length of
the circuit synthesis produced by Algorithm 1 is dependent on the choice $m$,
the size of the column sections. 
Here we have somewhat arbitrarily chosen $m=\mbox{round}((\log_2\,n)/2)$. The 
performance for some values of $n$ could be significantly improved by optimizing 
this choice. This would also smooth out the performance curve in 
Figure~\ref{fig:avg_length} for Algorithm 1.


\section{Conclusions and Future Work}
We have given an algorithm for linear reversible circuit synthesis
that is asymptotically optimal in the worst-case. We show that the
algorithm is also asymptotically faster than current methods. While
our results are primarily asymptotic, simulation results show that
even in the finite case our algorithm outperforms the current
synthesis method.  Applications of our work include circuit synthesis 
for quantum circuits.

While the primary motivations for the synthesis method we have given 
here are to provide an asymptotic bound on circuit complexity and a 
practical method to synthesize efficient circuits, another application 
is to bounds on circuit complexity for the finite case. In particular, 
we can use our method to determine an upper bound on the maximum 
number of gates required to synthesize any $n$ wire C-NOT circuit.
For this application the particular partitioning of the columns can
be very important. For example, much better bounds can be determined
if the size of the sections are a function of the location of the
section in the matrix. Sections to the left have more rows below the 
diagonal and therefore should be larger than sections towards the right 
of the matrix which have fewer rows below the diagonal. An ongoing
area of work is determining optimal column partitioning methods.

Our algorithm basically yields an efficient decomposition for matrices with 
elements in $\mathbb{F}_2$, and can be generalized in a straightforward 
manner for matrices over any finite field. The asymptotic 
size of the generalized decomposition is $O(n^2/\log_{|F|}n)$, where 
$|F|$ is the order of the finite field. Our algorithm, particularly
in this generalized form, is quite generic and may lend itself to a wide 
range of other applications. Related algorithms~\cite{arlazarov:oec:70} have
applications in finding the transitive closure of a graph, binary matrix
multiplication, and pattern matching.

A major area of future work is extending our results to more
general reversible circuits, particularly quantum circuits. Currently,
there is an asymptotic gap between the best upper and lower bounds on 
the worst-case circuit complexity both for general classical reversible 
circuits and quantum circuits. The gap for classical reversible circuits 
is the same logarithmic factor that previously existed for linear 
reversible circuits~\cite{shende:rlc:02}, which suggests it may be possible to extend our 
methods to this problem.  


\end{document}